\documentclass[11pt]{article}
\usepackage[centertags]{amsmath}
\usepackage{amsfonts}
\usepackage{amssymb}
\usepackage{amsthm}
\usepackage{newlfont}
\usepackage{graphicx}

\newfont{\bb}{msbm10 at 12pt}

\newtheorem{theorem}{Theorem}[section]

\newtheorem{remark}[theorem]{Remark}

\newtheorem{lemma}[theorem]{Lemma}

\begin{document}

\title{Concircular vector fields for Kantowski-Sachs and Bianchi type-III Spacetimes}
\author{ Suhail Khan$^{1^{*}}$, Amjad Mahmood$^{1}$ and  Ahmad T Ali$^{2,3}$ \\
1- Department of Mathematics, University of Peshawar, Peshawar,\\
Khyber Pakhtoonkhwa, Pakistan.\\
2- King Abdulaziz University, Faculty of Science, Department of Mathematics,\\
PO Box 80203, Jeddah, 21589, Saudi Arabia.\\
3- Mathematics Department, Faculty of Science, Al-Azhar University,\\
Nasr City, 11884, Cairo, Egypt.\\
$^*$E-mail: suhail{\_}74pk@yahoo.com}


\maketitle
\begin{abstract}
This paper intends to obtain concircular vector fields (CVFs) of Kantowski-Sachs and Bianch type-III spacetimes. For this purpose, ten conformal Killing equations and their general solution in the form of conformal Killing vector fields (CKVFs) are derived along with their conformal factors. The obtained conformal Killing vector fields are then placed in \textbf{Hessian} equations to obtain the final form of concircular vector fields. The existence of concircular symmetry imposes restrictions on the metric functions. The conditions imposing restrictions on these metric functions are obtained as a set of integrability conditions. It is shown that Kantowski-Sachs and Bianchi type-III spacetimes admit four, six, or fifteen dimensional concircular vector fields. It is established that for Einstein spaces, every conformal Killing vector field is a concircular vector field. Moreover, it is explored that every concircular vector field obtained here is also a conformal Ricci collineation.

\end{abstract}



\section{Introduction }
The theory of general relativity is governed by Einstein's field equations (EFEs), in which one side articulates the physics of spacetime and the other side describes geometry of the spacetime. EFEs are extremely nonlinear partial differential equations and it is very difficult to obtain their exact solutions. In order to solve EFEs, some symmetry restrictions are essential to apply on spacetimes.  Spacetime symmetries not only help us in finding  exact solutions of EFEs, but some of them  also provide invariant bases for the classification  of a spacetime \cite{Ali,Yadav,Attallah}. Spacetime classification, according to various types of symmetries is an important part of recent research in the field of general relativity \cite{hall1}. In the words of  G. S. Hall \cite{hall1}, a spacetime symmetry is a smooth vector field whose local flow preserves some geometrical feature of the spacetime, which refers to a specific tensor such as the energy momentum tensor, the metric tensor or to other aspects of the spacetimes such as geodesic structure. Some of the most basic symmetries of the spacetimes are Killing, homothetic and conformal vector fields.\\
\\
 Among other well-known spacetime symmetries, conformal symmetry is of specific interest. Mathematically specified a manifold $M$,  conformal symmetry is given by a vector field $X$ such that for a spacetime metric, along the integral curves created by $X$,  Lie derivative ($\mathcal{L}_{X}$) fulfills the relation,
\begin{equation}  \label{u2-b}
\mathcal{L}_X\,g_{ab}=g_{ab,c}\,X^{c} +g_{ac}\,X^c_{,b}+g_{bc}\,X^c_{,a}=2\,\alpha\,(x^m)g_{ab},
\end{equation}
where  $\alpha:M\rightarrow R $ is some smooth real valued function, called conformal factor. When $\alpha=0$, the conformal Killing vector fields become Killing vector fields $(KVFs)$ and when $\alpha_{,m}=0$, the solution of equation $(\ref{u2-b})$ is called homothetic vector fields $(HVFs)$. A conformal Killing vector field is known as proper conformal Killing vector field if $\alpha_{,m}\neq 0$ and is called special conformal Killing vector field if $\alpha_{m;n}=0\,\,\cite{hall1}$. In the case of Einstein spaces any conformal vector field  is also a Ricci collineation ($\mathcal{L}_{X}\,R_{ab}\,=\,0\,$) as well \cite{faridi1}. The importance and some applications of conformal Killing vector fields can be seen in $\cite{Tsamparlis,Maartens,Mason,Tupper}$.\\
\\
The conformal Ricci collineation  ($\mathcal{L}_{X}\,R_{ab}\,=\,2\,\alpha\,(x^m)R_{ab}$) were called concircular vector fields for Riemannian manifolds. Aaron Fialkow $\cite{Fialkow1}$, presented in his article the idea of concircular vector fields on a Riemannian manifold M. A vector field X is known as concircular vector field if the local flow generated by X consists of concircular mapping i.e. conformal mapping preserving geodesic circle. A transformation of the metric $g\rightarrow \tilde{g}=\,\frac{1}{\alpha^2}\,g$ is concircular if and only if
$\nabla^2\,\alpha=\,\frac{\Delta\,\alpha}{n}g$ $\cite{yano1, yano2, Tashiro1, Ferrand1}$, or equivalently the two Ricci tensors of $g$ and $\tilde{g}$ have the same traceless part $\cite{Kuhnal1}$.
Concircular vector fields have interesting applications in physics as well as in general relativity $\cite{Takeno}$, also it was proved by the author that a Lorentzian manifold is a generalized Robertson Walker spacetime if and only if it admits a timelike concircular vector field.

The notion of concircular transformations of Riemannian manifolds introduced by Yano  \cite{yano1, yano2} established the theory of concircular geometry. To explain his idea, let  $X^k$ is an infinitesimal concircular transformation (a vector field), if the infinitesimal point-transformation $\tilde{x}^k\,=\,x^k+\varepsilon\,X^k$ ($\varepsilon$ being an arbitrary infinitesimal constant) carries any geodesic circle into another one. Necessary and sufficient conditions for $X^k$ to be a concircular vector field is given by Ishihara $\cite{ishi1}$, which is given as follows:
\begin{theorem} \label{tm-1}
A vector field $X$ is said to be a concircular vector field ( that is, $X^k$ to be an infinitesimal concircular transformation), if it satisfies the following necessary and sufficient conditions
\begin{equation}\label{u21-a}
\mathcal{L}_{X}\,g_{ab}\,=\,g_{ab,c}\,X^c+g_{bc}\,X^c_{\,\,,a}+g_{ac}\,X^{c}_{\,\,,b}\,=\,2\,\alpha\,g_{ab},
\end{equation}
\begin{equation}\label{u21-b}
\nabla_a\,\nabla_b\,\psi\,=\,\psi_{,ab}-\Gamma_{ab}^{\,c}\,\psi_{,c}\,=\,\beta\,g_{ab},
\end{equation}
where $\Gamma_{ab}^{\,c}$ indicates the second kind Christoffel symbols,  $\nabla_a\,\nabla_b$ is the \textbf{Hessian} while  $\alpha$ and $\beta$ are conformal factors.
\end{theorem}
Basically $\alpha$ is the factor of dilation of the infinitesimal concircular transformation $X^k$. The above theorem shows that the vector fields considered in \cite{faridi1} are the particular form of concircular vector fields. One can obtain conformal Killing vector fields when equation (\ref{u21-a}) is satisfied and  a concircular vector field when equation (\ref{u21-b}) is satisfied. Taking the covariant derivative on both sides of equation (\ref{u21-b}) and taking account of the Ricci identity, the following relation can be established easily
\begin{equation}\label{u21-b-2}
R_{cba}^{\,\,\,\,\,\,\,l}\,\alpha_{,l}\,=\,g_{cb}\,\beta_{,a}-g_{ac}\,\beta_{,b},
\end{equation}
$R_{cba}^{\,\,\,\,\,\,l}$ being the Riemann curvature tensor of the manifold and  $\beta_{,a}\,=\,\partial_a\beta$.\\
Also, $\alpha\,=\,\dfrac{\mathbf{div}\,X}{n}$ $\cite{tsam1}$, that is  the conformal factor is equal to the constant times divergence of $X$. Kuhnel and Rademacher $\cite{kuhnel1}$ proved that for a general n-dimensional manifold admitting CKVFs, their special conformal Killing vector field  ($\nabla_a\,\nabla_b\,\alpha\,=\,0\,$) is equivalent to its Ricci collineation.
The Ricci tensor of Einstein spaces remains proportional to its metric, i.e., $R_{ab}\,=\,K\,g_{ab}$, where $K$ is the constant of proportionality.  For Einstein's spaces, $K$ and $R$ are related as $R\,=\,n\,K$ \cite{berg1}.
\begin{lemma} \label{lm-2} $\cite{ishi1}$: In Einstein spaces every conformal Killing vector field is a concircular vector field.
\end{lemma}
Also a relation between $\alpha$, and $\beta$ for manifolds of constant curvature becomes:
\begin{equation}  \label{u31-01}
\beta\,=\,\dfrac{R\,\alpha}{n\,(n-1)}.
\end{equation}
\section{Conformal and Concircular Equations}

Consider a spacetime metric in the  spherical coordinate system $(t,r,\theta,\phi)$ with the line element $\cite{steph1}$:
\begin{equation}  \label{u31}
ds^2=-dt^2+A^2(t)\,dr^2+B^2(t)\left(d\theta^2+h^2(\theta)d\phi^2\right),
\end{equation}
where $A$ and $B$ are no where zero functions of $t$ only such that for $h(\theta)=\sin(\theta)$, spacetime (\ref{u31}) becomes Kantowski-Sachs spacetime and for $h(\theta)=\sinh(\theta)$, spacetime (\ref{u31}) becomes Bianchi type-III spacetime. These spacetimes admit minimum four Killing vector fields $\dfrac{\partial}{\partial\,r}$, $\dfrac{\partial}{\partial\,\phi}$, $\cos[\phi]\,\dfrac{\partial}{\partial\,\theta}-\,\dfrac{h'(\theta)}{h(\theta)}\sin[\phi]\,\dfrac{\partial}{\partial\,\phi}$ and $\sin[\phi]\,\dfrac{\partial}{\partial\,\theta}+\,\dfrac{h'(\theta)}{h(\theta)}\cos[\phi]\,\dfrac{\partial}{\partial\,\phi}$ \cite{kantowski}.
Using the metric components above in equations (\ref{u21-a}) and (\ref{u21-b}), the following two systems of partial differential equations are obtained as:
\begin{equation}  \label{u33-1}
X^0_{\,,0}\,=\,\alpha,
\end{equation}
\begin{equation}  \label{u33-2}
A^{2}\,X^{1}_{\,,0}-X^0_{\,,1}\,=\,0,
\end{equation}
\begin{equation}  \label{u33-3}
B^{2}\,X^{2}_{\,,0}-X^0_{\,,2}\,=\,0,
\end{equation}
\begin{equation}  \label{u33-4}
B^{2}\,h^2(\theta)\,X^{3}_{\,,0}-X^0_{\,,3}\,=\,0,
\end{equation}
\begin{equation}  \label{u33-5}
A\,X^{1}_{\,,1}+A'\,X^0\,=\,\alpha\,A,
\end{equation}
\begin{equation}  \label{u33-6}
A^2\,X^{1}_{\,,2}+B^{2}\,X^2_{\,,1}\,=\,0,
\end{equation}
\begin{equation}  \label{u33-7}
A^2\,X^{1}_{\,,3}+B^{2}\,h^2(\theta)\,X^3_{\,,1}\,=\,0,
\end{equation}
\begin{equation}  \label{u33-8}
B\,X^2_{\,,2}+B'\,X^0\,=\,\alpha\,B,
\end{equation}
\begin{equation}  \label{u33-9}
X^{2}_{\,,3}+h^2(\theta)\,X^3_{\,,2}\,=\,0,
\end{equation}
\begin{equation}  \label{u33-10}
B\,\left(\dfrac{h'(\theta)}{h(\theta)}\,X^{2}+X^3_{\,,3}\right)+B'\,X^0\,=\,\alpha\,B,
\end{equation}
and
\begin{equation}  \label{u33-11}
\psi_{\,,tt}+\phi=\,0,
\end{equation}
\begin{equation}  \label{u33-12}
\psi_{\,,rr}-A(t)\,\left(A'(t)\,\psi_{\,,t}+A(t)\,\phi\right)=\,0,
\end{equation}
\begin{equation}  \label{u33-13}
\psi_{\,,\theta\,\theta}-B(t)\,\left(B'(t)\,\psi_{\,,t}+B(t)\,\phi\right)=\,0,
\end{equation}
\begin{equation}  \label{u33-14}
\psi_{\,,\phi\,\phi}-{B(t)}{h^2(\theta)}\,\left(B'(t)\,\psi_{\,,t}+B(t)\,\phi\right)+h(\theta)\,h'(\theta)\,\psi_{\,,\theta}=\,0,
\end{equation}
\begin{equation}  \label{u33-15}
\psi_{\,,tr}-\dfrac{A'(t)}{A(t)}\,\psi_{\,,r}=\,0,
\end{equation}
\begin{equation}  \label{u33-16}
\psi_{\,,t\theta}-\dfrac{B'(t)}{B(t)}\,\psi_{\,,\theta}=\,0,
\end{equation}
\begin{equation}  \label{u33-17}
\psi_{\,,t\phi}-\dfrac{B'(t)}{B(t)}\,\psi_{\,,\phi}=\,0,
\end{equation}
\begin{equation}  \label{u33-18}
\psi_{\,,r\,\theta}=\,0,
\end{equation}
\begin{equation}  \label{u33-19}
\psi_{\,,r\,\phi}=\,0,
\end{equation}
\begin{equation}  \label{u33-20}
\psi_{\,,\theta\,\phi}-\dfrac{h'(\theta)}{h(\theta)}\,\psi_{\,,\phi}=\,0.
\end{equation}

We first solve Eqs (\ref{u33-1})--(\ref{u33-10}) by direct integration technique to find the components $X^0$, $X^1$, $X^2$ and
$X^3$ of the Conformal Killing Vector Fields. We will then substitute the resulting CKVF components in the Eqs (\ref{u33-11})--(\ref{u33-20}) to obtain differential constraints, which when solved will give us the Concircular vector fields. The process of obtaining these components is explained as follows:\\
First differentiating Eq (\ref{u33-2}) with respect to $\theta$, Eqs. (\ref{u33-3}) and (\ref{u33-4}) with respect to $r$ and Eqs. (\ref{u33-6}) and (\ref{u33-7}) with respect to $t$ after dividing them by $A^2$, we obtain the following relations:
\begin{equation}  \label{u33-21}
\left(\dfrac{B\,X^2_{,r}}{A}\right)_{,t}\,=\,\left(\dfrac{B\,X^3_{,r}}{A}\right)_{,t}\,=\,0.
\end{equation}
Integrating the above equation, substituting the results in Eqs. (\ref{u33-2})-(\ref{u33-10}) and solving Eqs. (\ref{u33-3}) and (\ref{u33-6}), we get
\begin{equation}  \label{u33-22}
\left\{
  \begin{array}{ll}
    X^0\left(t,r,\theta,\phi\right)\,=\,X^{00}+\left(B\,A'-A\,B'\right)\,\left(X^{20}+X^{21}\right)-A\,B\,X^{21}_{,t},\\
    \\
    X^1\left(t,r,\theta,\phi\right)\,=\,\dfrac{B}{A}\,\left(X^{10}-X^{20}_{,t}\right),\\
\\
X^2\left(t,r,\theta,\phi\right)\,=\,\dfrac{A}{B}\,\left(X^{20}+X^{21}\right)_{,\theta},\\
    \\
X^3\left(t,r,\theta,\phi\right)\,=\,\dfrac{A}{B}\,\left(X^{30}+X^{31}\right)_{,\phi},
  \end{array}
\right.
\end{equation}
where $X^{00},\,X^{10}\,=\,X^{00},\,X^{10}\left(t,r,\phi\right)$, $X^{20},\,X^{30}\,=\,X^{20},\,X^{30}\left(r,\theta,\phi\right)$ and $X^{21},\,X^{31}\,=\,X^{21},\,X^{31}\left(t,\theta,\phi\right)$. Again, substituting the results above in Eqs. (\ref{u33-2}), (\ref{u33-4}), (\ref{u33-5}), (\ref{u33-7})-(\ref{u33-10}) and then solving  Eqs. (\ref{u33-2}),  (\ref{u33-4}) and (\ref{u33-7}), we get:
\begin{equation}  \label{u33-23}
\left\{
  \begin{array}{ll}
    X^{00}\,=\,B^2\,Y^{00}_{,t}+\left(B\,A'-A\,B'\right)\,\left(Y^{10}+Y^{12}\right)-A\,B\,Y^{10}_{,t},\\
    \\
    X^{10}\left(t,r,\phi\right)\,=\,\left(Y^{10}+Y^{12}\right)_{,r},\\
    \\
    X^{20}\left(r,\theta,\phi\right)\,=\,Y^{12}-h(\theta)\,\left(Y^{201}+Y^{202}+Y^{203}\right),\\
    \\
     X^{21}\left(t,\theta,\phi\right)\,=\,\dfrac{B}{A}\,\left(Y^{213}+Y^{214}-Y^{00}\right)+h^2(\theta)\,\left(Y^{30}+X^{31}\right),\\
    \\
    X^{30}\left(r,\theta,\phi\right)\,=\,Y^{30}_{\phi}+\dfrac{1}{h^2(\theta)}\,\left(X^{20}_{\phi}-Y^{12}\right).
  \end{array}
\right.
\end{equation}
where $Y^{00}\,=\,Y^{00}\left(t,\phi\right)$, $Y^{10}\,=\,Y^{10}\left(t,r\right)$,  $Y^{12}\,=\,Y^{12}\left(r,\phi\right)$, $Y^{201}\,=\,Y^{201}\left(\theta,\phi\right)$, $Y^{202}\,=\,Y^{202}\left(r,\theta\right)$, $Y^{203}\,=\,Y^{203}\left(r,\phi\right)$, $Y^{213}\,=\,Y^{213}\left(t,\theta\right)$, $Y^{214}\,=\,Y^{214}\left(\theta,\phi\right)$ and $Y^{30}\,=\,Y^{30}\left(\theta,\phi\right)$. Now, with the help of above information we can solve Eq. (\ref{u33-9}) to get the general form of conformal Killing vector fields as follows:
\begin{equation}  \label{0u41}
\left\{
  \begin{array}{ll}
    X^0\,=\,B\,F_0+B^2\,\Big[H_1\,h'(\theta)+h(\theta)\Big(H_2\,\sin[\phi]+H_3\,\cos[\phi]\Big)\Big]_{,t},\\
\\
X^1\,=\,\left[F_4+F_0\,\int\dfrac{B}{A^2}\,dt-\dfrac{B}{A}\,\Big(h'(\theta)\,F_1+h(\theta)\Big[F_2\,\sin[\phi]+F_3\,\cos[\phi]\Big]\Big)\right]_{,r},\\
\\
X^2\,=\,\Big(c_3+H_3\,h'(\theta)\Big)\,\cos[\phi]+\Big(c_2+H_2\,h'(\theta)\Big)\,\sin[\phi]-H_1\,h(\theta),\\
\\
X^3\,=\,c_1+\dfrac{h'(\theta)}{h(\theta)}\,\Big(c_2\,\cos[\phi]+c_3\,\sin[\phi]\Big)+\dfrac{1}{h(\theta)}\,\Big(H_2\,\cos[\phi]+H_3\,\sin[\phi]\Big),
  \end{array}
\right.
\end{equation}
where $c_1$, $c_2$ and $c_3$ are arbitrary constants, $H_j\,=\,H_j(t,r)\,=\,G_j(t)+\dfrac{A(t)\,F_j(r)}{B(t)}$, such that $G_{j}\,=G_{j}(t)$ are functions of $t$ only for $j\,=\,1,2,3$ while $F_{i}\,=F_{i}(r)$ are functions of $r$ only for $i\,=\,0,1,2,3,4$ and the conformal factor is given by $\alpha\,=\,X^0_{,t}$. The above system constitutes a general solution of the Conformal Killing Vector Field Equations (\ref{u33-1})-(\ref{u33-10}).
Substituting these values of $X^0$, $X^1$, $X^2$ and $X^3$ in the remaining Eqs. (\ref{u33-5}), (\ref{u33-8}) and (\ref{u33-10}), expanding it with the aid of \textit{mathematica program} and set the coefficients involving $\sin[\phi]$, $\cos[\phi]$, $h(\theta)$ and $h'(\theta)$ and various products equal zero, give the following set of over-determined integrability differential constraints of Conformal Killing Vector Fields:

\begin{equation}  \label{0u33-1}
\ddot{F}_4+\ddot{F}_0\,\left(\int\dfrac{B}{A^2}\,dt\right)\,=\,A\,\left(\dfrac{B}{A}\right)'\,F_0,
\end{equation}

\begin{equation}  \label{0u33-2}
\dfrac{\ddot{F}_j}{A}-\dfrac{A\,F_j}{B}\,\Bigg[A\,\left(\dfrac{B'}{A}\right)'-B\,\left(\dfrac{A'}{A}\right)'\Bigg]
\,=\,A\,\left(\dfrac{B}{A}\right)'\,G'_j+\Big(B\,G'_j\Big)',\,\,\,\,\,\,\,j\,=\,1,2,3,
\end{equation}

\begin{equation}  \label{0u33-3}
F_j\,\Bigg[B\,\left(\dfrac{A'}{B}\right)'-A\,\left(\dfrac{B'}{B}\right)'+\dfrac{A}{B^2}\Bigg]\,=\,\dfrac{G_j}{B}+\Big(B\,G'_j\Big)',\,\,\,\,\,\,\,j\,=\,1,2,3,
\end{equation}
where $\dot{F}_i\,=\,\dfrac{dF_i}{dr}$ and $G'_j\,=\,\dfrac{dG_j}{dt}$.

Also, if we put the values of $X^0$, $X^1$, $X^2$ and $X^3$ in the Eqs. (\ref{u33-11})--(\ref{u33-20}) and set the coefficients involving $\sin[\phi]$, $\cos[\phi]$, $h(\theta)$ and $h'(\theta)$ and various products equal zero, we get the following set of over-determined integrability differential constraints of concircular vector fields:

\begin{equation}  \label{2u33-1}
\left(\dfrac{B'}{A}\right)'\,\dot{F}_0\,=\,0,
\end{equation}
\begin{equation}  \label{2u33-2}
\left(\dfrac{B}{A}\right)'\,\left(\dfrac{B'}{A}\right)'\,\dot{F}_j\,=\,0,\,\,\,\,\,j\,=\,1,2,3,
\end{equation}

\begin{equation}  \label{3-4u33-1}
\Bigg[B\,\left(\dfrac{A'}{B}\right)'+A\,\left(\dfrac{B'}{B}\right)'-\dfrac{A}{B^2}\Bigg]\,H'_j\,=\,0,\,\,\,\,\,j\,=\,1,2,3,
\end{equation}

\begin{equation}  \label{5u33-1}
\Big[A^2\,\left(2\,\mu_1\,B'+B\,\mu'_1\right)+B\,\mu'_2\Big]\,F_0+2\,\mu_2\,\left[\ddot{F}_4+\ddot{F}_0\,\left(\int\dfrac{B}{A^2}\,dt\right)\right]\,=\,0,
\end{equation}
\begin{equation}  \label{5u33-2}
  \begin{array}{ll}
2\,B\,\mu_1\,\ddot{F}_j-A\,F_j\,\Bigg[2\,\mu_1\,\left(B\,A''-A\,B''\right)-\left(\dfrac{A}{B}\right)'\,\left(A^2\,\mu'_1+\mu'_2\right)\Bigg]\\
\\
\,\,\,\,\,\,\,\,\,\,\,\,\,\,\,\,\,
\,=\,A\,B\,\Big[B\,G'_j\,\left(A^2\,\mu'_1+\mu'_2\right)+2\,A^2\,\mu_1\,\left(2\,B'\,G'_j+B\,G''_j\right)\Big],\,\,\,\,\,j\,=\,1,2,3,
  \end{array}
\end{equation}

\begin{equation}  \label{6-7u33-1}
\Big(B^2\,\mu_2-A^2\,\mu_3\Big)\,\dot{F}_j\,=\,0,\,\,\,\,\,j\,=\,1,2,3,
\end{equation}

\begin{equation}  \label{8-10u33-1}
\left[2\,B\,B'\,\mu_1+B^2\,\mu'_1+\mu'_3\right]\,F_0=\,0,
\end{equation}
\begin{equation}  \label{8-10u33-2}
  \begin{array}{ll}
   B^2\,\left(B^2\,\mu'_1+\mu'_3\right)\,H'_j-2\,\mu_3\,H_j-2\,B^2\,\mu_1\,\Big[\left(B\,A''-A\,B''\right)\,F_j \\
   \\
   -B^2\,\left(B\,G''_j+2\,B'\,G'_j\right)\Big]\,=\,0,\,\,\,\,\,j\,=\,1,2,3,
  \end{array}
\end{equation}
where $\beta\,=\,-\mu_1\,X^{0}_{\,,0}-\dfrac{\mu'_{1}\,X^0}{2}$.

From the conditions (\ref{0u33-1})--(\ref{0u33-3}), we can study all possible cases when the spacetime metric (\ref{u31}) admit Conformal Killing Vector Fields. To avoid lengthy details we shall write only the final results. The conclusion of the final results can be written in the following as different
possibilities:

\section{Conformal Killing Vector Fields}

If we differentiate Eq. (\ref{0u33-3}) with respect to $r$, we find the following condition:
\begin{equation}  \label{0u33-30}
\dot{F}_j\,\Bigg[B\,\left(\dfrac{A'}{B}\right)'-A\,\left(\dfrac{B'}{B}\right)'+\dfrac{A}{B^2}\Bigg]\,=\,0,\,\,\,\,\,\,\,j\,=\,1,2,3,
\end{equation}
The above condition leads to two cases:

\textbf{(I):} $B\,\left(\dfrac{A'}{B}\right)'-A\,\left(\dfrac{B'}{B}\right)'+\dfrac{A}{B^2}\,=\,0$.

\textbf{(II):} $B\,\left(\dfrac{A'}{B}\right)'-A\,\left(\dfrac{B'}{B}\right)'+\dfrac{A}{B^2}\,\neq\,0$ and $\dot{F}_1\,=\,\dot{F}_2\,=\,\dot{F}_3\,=\,0$.

Now, we will discuss all possibilities and hence find the following solutions:

\textbf{Conformal Killing  Vector Field (1):}  In this case, the general solution of equation \textbf{(I)} takes the following form:

$B(t)\,=\,\dfrac{A(t)}{\gamma_0}\,\cosh\left[\Psi\right]$, where $\gamma_0$ is an arbitrary non zero constant, while $\Psi\,=\,\gamma_0\,\int\dfrac{dt}{A(t)}$ such that $A(t)$ is an arbitrary no where zero function of $t$ only. Therefore, we find

\begin{equation}  \label{u41-00}
\left\{
  \begin{array}{ll}
    G_1(t)\,=\,\gamma_0\,\left(a_3\,\mathrm{tanh}[\Psi]-c_1\,\mathrm{sech}[\Psi]\right),
    G_2(t)\,=\,\gamma_0\,\left(a_7\,\mathrm{tanh}[\Psi]-c_2\,\mathrm{sech}[\Psi]\right),\\
    \\
G_3(t)\,=\,\gamma_0\,\left(a_6\,\mathrm{tanh}[\Psi]-c_3\,\mathrm{sech}[\Psi]\right),
F_0(r)\,=\,\gamma_0\,\left(a_1\,\mathrm{e}^{\gamma_0\,r}+a_2\,\mathrm{e}^{-\gamma_0\,r}\right),\\
    \\
F_1(r)\,=\,c_1-a_4\,\mathrm{e}^{\gamma_0\,r}-a_5\,\mathrm{e}^{-\gamma_0\,r},
F_2(r)\,=\,c_2+a_{10}\,\mathrm{e}^{\gamma_0\,r}-a_{11}\,\mathrm{e}^{-\gamma_0\,r},\\
    \\
F_3(r)\,=\,c_3+a_8\,\mathrm{e}^{\gamma_0\,r}-a_9\,\mathrm{e}^{-\gamma_0\,r},
F_4(r)\,=\,c_4+a_{12}\,r,
  \end{array}
\right.
\end{equation}
where $\gamma_0$, $c_j$ and $a_i$ are arbitrary constants, $i=1,...,12$,  $j=1,...,4$.
The components of conformal Killing vector fields are obtained as:
\begin{equation}  \label{u41-CKVF-1}
\left\{
  \begin{array}{ll}
X^0\,=\,\Bigg[\left(a_1\,\mathrm{e}^{\gamma_0\,r}+a_2\,\mathrm{e}^{-\gamma_0\,r}\right)\,\cosh\left[\Psi\right]
+h'(\theta)\Big[a_3+\left(a_4\,\mathrm{e}^{\gamma_0\,r}+a_5\,\mathrm{e}^{-\gamma_0\,r}\right)\,\sinh\left[\Psi\right]\Big]\\
\\
+h(\theta)\Bigg(a_6\,\cos[\phi]+a_7\,\sin[\phi]+\Big[\left(a_8\,\mathrm{e}^{\gamma_0\,r}+a_9\,\mathrm{e}^{-\gamma_0\,r}\right)\,\cos[\phi]\\
\\
+\left(a_{10}\,\mathrm{e}^{\gamma_0\,r}+a_{11}\,\mathrm{e}^{-\gamma_0\,r}\right)\,\sin[\phi]\Big]\,\sinh\left[\Psi\right]\Bigg)\Bigg]\,A(t),\\
\\
X^1\,=\,a_{12}+\left(a_1\,\mathrm{e}^{\gamma_0\,r}-a_2\,\mathrm{e}^{-\gamma_0\,r}\right)\,\sinh\left[\Psi\right]
+\Bigg(h'(\theta)\left(a_4\,\mathrm{e}^{\gamma_0\,r}-a_5\,\mathrm{e}^{-\gamma_0\,r}\right)\\
\\
-h(\theta)\Bigg[\left(a_8\,\mathrm{e}^{\gamma_0\,r}+a_9\,\mathrm{e}^{-\gamma_0\,r}\right)\,\cos[\phi]
+\left(a_{10}\,\mathrm{e}^{\gamma_0\,r}+a_{11}\,\mathrm{e}^{-\gamma_0\,r}\right)\,\sin[\phi]\Bigg]\Bigg)\,\cosh\left[\Psi\right],\\
\\
X^2\,=a_{13}\,\cos[\phi]+a_{14}\,\sin[\phi]+\gamma_0h'(\theta)\,\Bigg[\left(a_{6}\,\cos[\phi]+a_{7}\,\sin[\phi]\right)\tanh[\Psi]\\
\\
+\Big[\left(a_8\,\mathrm{e}^{\gamma_0\,r}-a_9\,\mathrm{e}^{-\gamma_0\,r}\right)\,\cos[\phi] +\left(a_{10}\,\mathrm{e}^{\gamma_0\,r}-a_{11}\,\mathrm{e}^{-\gamma_0\,r}\right)\,\sin[\phi]\Big]\,\mathrm{sech}[\Psi]\Bigg]\\
\\
-\gamma_0\,h(\theta)\,\Big[a_3\,\tanh[\Psi]-\left(a_4\,\mathrm{e}^{\gamma_0\,r}+a_5\,\mathrm{e}^{-\gamma_0\,r}\right)\,\mathrm{sech}\left[\Psi\right]\Big],\\
\\
X^3\,=\,a_{15}+\dfrac{h'(\theta)}{h(\theta)}\,\left(a_{14}\,\cos[\phi]-a_{13}\,\sin[\phi]\right)
+\gamma_0\,\dfrac{1}{h(\theta)}\,\Bigg[\left(a_{7}\,\cos[\phi]-a_{6}\,\sin[\phi]\right)\tanh[\Psi]\\
\\
+\Bigg(\left(a_{10}\,\mathrm{e}^{\gamma_0\,r}-a_{11}\,\mathrm{e}^{-\gamma_0\,r}\right)\,\cos[\phi] -\left(a_{8}\,\mathrm{e}^{\gamma_0\,r}-a_{9}\,\mathrm{e}^{-\gamma_0\,r}\right)\,\sin[\phi]\Bigg)\,\mathrm{sech}[\Psi]\Bigg],
  \end{array}
\right.
\end{equation}
where $A(t)$ is an arbitrary no where zero function of $t$ while $a_i$, $i=1,...,15$ are arbitrary constants.
The conformal factor takes the following form
\begin{equation}  \label{u41-CF-I}
  \begin{array}{ll}
\alpha(t,x,y,z)\,=\,\Big(\gamma_0\,\sinh\left[\Psi\right]+A'(t)\,\cosh\left[\Psi\right]\Big)\,\left(a_1\,\mathrm{e}^{\gamma_0\,r}+a_2\,\mathrm{e}^{-\gamma_0\,r}\right)\\
    \\
    +A'(t)\,\Big[a_3\,h'(\theta)+\left(a_{6}\,\cos[\phi]+a_{7}\,\sin[\phi]\right)\,h(\theta)\Big]+\Big(\gamma_0\,\cosh\left[\Psi\right]\\
    \\
    +A'(t)\,\sinh\left[\Psi\right]\Big)\,\Bigg(\left(a_4\,\mathrm{e}^{\gamma_0\,r}+a_5\,\mathrm{e}^{-\gamma_0\,r}\right)\,h'(\theta)\\
    \\
    +\Big[\left(a_8\,\mathrm{e}^{\gamma_0\,r}+a_9\,\mathrm{e}^{-\gamma_0\,r}\right)\,\cos[\phi]
    +\left(a_{10}\,\mathrm{e}^{\gamma_0\,r}+a_{11}\,\mathrm{e}^{-\gamma_0\,r}\right)\,\sin[\phi]\Big]\,\sin[\theta]\Bigg).
  \end{array}
\end{equation}
The final form of the conformal Killing vector fields admit the 15-dimensional Lie algebra of the metric:
\begin{equation}  \label{u41-M-I}
ds^2=-dt^2+A^2(t)\,\Bigg[dr^2+\gamma_0^{-2}\,\cosh^2\left[\gamma_0\,\int\,A^{-1}(t)\,dt\right]\,\Big(d\theta^2+h^2(\theta)\,d\phi^2\Big)\Bigg],
\end{equation}
where  $\gamma_0$ is an arbitrary non zero constant.


\begin{remark} \label{rm-1-0}
It is important to note that the conformal Killing vector fields (\ref{u41-CKVF-1}) do not admit proper homothetic vector fields when $a_1\,=\,a_2\,=\,a_3\,=\,a_6\,=\,a_7\,=\,0$, $A(t)\,=\,\dfrac{\gamma_0\,\sinh[\delta_0\,t]}{\delta_0}$ and $B(t)\,=\,\dfrac{\cosh[\delta_0\,t]}{\delta_0}$, where $\gamma_0$ and $\delta_0$ are arbitrary constants. Thus, Killing vector fields are obtained after subtracting proper conformal vector fields as follows:
\end{remark}
\begin{equation}  \label{u41-CVF-1A-00}
\left\{
  \begin{array}{ll}
    X^0\,=\,\gamma_0\,\Bigg[h'(\theta)\Big[c_1\,\mathrm{e}^{\gamma_0\,r}+c_2\,\mathrm{e}^{-\gamma_0\,r}\Big]
+h(\theta)\Bigg(\Big[c_3\,\mathrm{e}^{\gamma_0\,r}+c_4\,\mathrm{e}^{-\gamma_0\,r}\Big]\,\cos[\phi]\\
\\
\,\,\,\,\,\,\,\,\,\,\,\,\,\,\,\,\,\,\,\,\,\,\,\,\,\,\,\,\,\,\,\,\,\,\,\,\,\,\,\,
\,\,\,\,\,\,\,\,\,\,\,\,\,\,\,\,\,\,\,\,\,\,\,\,\,\,\,\,\,\,\,\,\,\,\,\,\,\,\,\,
+\Big[c_{5}\,\mathrm{e}^{\gamma_0\,r}+c_{6}\,\mathrm{e}^{-\gamma_0\,r}\Big]\,\sin[\phi]\Bigg)\Bigg],\\
\\
X^1\,=\,a_{7}-\delta_0\,\,\coth\left[\delta_0\,t\right]\Bigg(\cos[\theta]\left(c_1\,\mathrm{e}^{\gamma_0\,r}-c_2\,\mathrm{e}^{-\gamma_0\,r}\right)\\
\\
+h(\theta)\Big[\left(c_3\,\mathrm{e}^{\gamma_0\,r}-c_4\,\mathrm{e}^{-\gamma_0\,r}\right)\,\cos[\phi]
+\left(c_{5}\,\mathrm{e}^{\gamma_0\,r}-c_{6}\,\mathrm{e}^{-\gamma_0\,r}\right)\,\sin[\phi]\Big]\Bigg),\\
\\
X^2\,=\,c_{8}\,\cos[\phi]+c_{9}\,\sin[\phi]-\gamma_0\,\delta_0\,\,\mathrm{tanh}[\delta_0\,t]\,
\Bigg(h(\theta)\,\left(c_1\,\mathrm{e}^{\gamma_0\,r}+c_2\,\mathrm{e}^{-\gamma_0\,r}\right)\\
\\
+h'(\theta)\,\Big[\left(c_3\,\mathrm{e}^{\gamma_0\,r}+c_4\,\mathrm{e}^{-\gamma_0\,r}\right)\,\cos[\phi] +\left(c_{5}\,\mathrm{e}^{\gamma_0\,r}+c_{6}\,\mathrm{e}^{-\gamma_0\,r}\right)\,\sin[\phi]\Big]\Bigg),\\
\\
X^3\,=\,c_{10}+\dfrac{h'(\theta)}{h(\theta)}\,\left(c_{9}\,\cos[\phi]-c_{8}\,\sin[\phi]\right)+\gamma_0\,\delta_0\,\csc[\theta]\,\,\mathrm{tanh}[\delta_0\,t]\,\Bigg[\\
\\
\,\,\,\,\,\,\,\,\,\,\,\,\,\,\,\,\,\,\,\,\,\,\,\,\,\,\,\,\,\,\,\,\,\,\,\,\,\,\,\,
\left(c_{5}\,\mathrm{e}^{\gamma_0\,r}+c_{6}\,\mathrm{e}^{-\gamma_0\,r}\right)\,\cos[\phi]-\left(c_{3}\,\mathrm{e}^{\gamma_0\,r}+c_{4}\,\mathrm{e}^{-\gamma_0\,r}\right)\,\sin[\phi]\Bigg],
  \end{array}
\right.
\end{equation}
where $c_i$, $i=1,...,10$ are arbitrary constants.\\

\textbf{Conformal Killing  Vector Field (2):} In the case  \textbf{(II)}, $B(t)\,=\,\gamma_0\,A(t)$, where $\gamma_0$ is arbitrary non zero constant and $A(t)$ is an arbitrary no where zero function of $t$ only. The components of conformal Killing vector fields now become:
\begin{equation}  \label{u41-CKVF-4}
\left\{
  \begin{array}{ll}
    X^0\,=\,A(t)\,\left(a_1+a_2\,r\right),\\
\\
X^1\,=\,a_{3}+a_2\,\int\dfrac{dt}{A(t)},\\
\\
X^2\,=a_{4}\,\cos[\phi]+a_{5}\,\sin[\phi],\\
\\
X^3\,=\,a_{6}-\dfrac{h'(\theta)}{h(\theta)}\,\left(a_{4}\,\sin[\phi]-a_{5}\,\cos[\phi]\right),
  \end{array}
\right.
\end{equation}
where  $a_i$, $i=1,...,6$ are arbitrary constants. The conformal factor takes the following form
\begin{equation}  \label{u41-CF-4}
  \begin{array}{ll}
    \alpha(t,x,y,z)\,=\,A'(t)\,\left(a_1+a_2\,r\right).
  \end{array}
\end{equation}
The final form of the conformal Killing vector fields admit six-dimensional Lie algebra of the metric:
\begin{equation}  \label{u41-M-4}
ds^2=-dt^2+A^2(t)\,\Big[dr^2+\gamma_0^2\,\Big(d\theta^2+h^2(\theta)\,d\phi^2\Big)\Big],
\end{equation}
where $A(t)$ and $\gamma_0$ are as defined above.
\begin{remark} \label{rm-2-0}
The conformal Killing vector fields (\ref{u41-CKVF-4}) admit homothetic vector fields when $a_2\,=\,0$, $A(t)\,=\,t+c_2$ and $B(t)\,=\,\gamma_0\,\left(t+c_2\right)$, where $\gamma_0$ and $c_2$ are arbitrary constants, such that $\gamma_0\neq0$. Five dimensional homothetic vector fields after subtracting one proper conformal vector field, are obtained as follows:
\end{remark}
\begin{equation}  \label{u41-CKVF-4-00}
\left\{
  \begin{array}{ll}
    X^0\,=\,c_1\,t+c_2,\\
\\
X^1\,=\,c_{3},\\
\\
X^2\,=c_{4}\,\cos[\phi]+c_{5}\,\sin[\phi],\\
\\
X^3\,=\,c_{6}-\dfrac{h'(\theta)}{h(\theta)}\,\left(c_{4}\,\sin[\phi]-c_{5}\,\cos[\phi]\right),
  \end{array}
\right.
\end{equation}
where $c_i$, $i=1,...,6$ are arbitrary constants. The conformal factor takes the following form \\
\\
$\alpha(t,x,y,z)\,=\,c_1$. Note that the above homothetic vector fields are the same as obtained earlier by G. Shabbir and F. Iqbal $\cite{shabbir1}$.

\textbf{Conformal Killing  Vector Field (3):} In this case, solution of equations (\ref{0u33-1})-(\ref{0u33-3}), exist when $B(t)\,=\,A(t)\,\left(\gamma_1\,\sin[\Psi]+\gamma_2\,\cos[\Psi]\right)$, such that
$\Psi\,=\,\gamma_0\,\int\dfrac{dt}{A(t)}$, where $\gamma_0$, $\gamma_1$, $\gamma_2\neq0$ are arbitrary constants while $A(t)$ is an arbitrary no where zero function of $t$. The components of conformal Killing vector fields in this case become:
\begin{equation}  \label{u41-CKVF-2}
\left\{
  \begin{array}{ll}
X^0\,=\,A(t)\,\left(\gamma_1\,\sin[\Psi]+\gamma_2\,\cos[\Psi]\right)\,\left(a_1\,\sin[\gamma_0\,r]+a_2\,\cos[\gamma_0\,r]\right),\\
\\
X^1\,=\,a_{3}-\left(\gamma_1\,\cos[\Psi]-\gamma_2\,\sin[\Psi]\right)\left(a_1\,\cos[\gamma_0\,r]-a_2\,\sin[\gamma_0\,r]\right),\\
\\
X^2\,=a_{4}\,\cos[\phi]+a_{5}\,\sin[\phi],\\
\\
X^3\,=\,a_{6}-\dfrac{h'(\theta)}{h(\theta)}\,\left(a_{4}\,\sin[\phi]-a_{5}\,\cos[\phi]\right),
  \end{array}
\right.
\end{equation}
where $a_i$, $i=1,...,6$ are arbitrary constants. Also the conformal factor takes the following form
\begin{equation}  \label{u41-CF-2}
  \begin{array}{ll}
    \alpha(t,x,y,z)\,=\,\Bigg[\gamma_0\,\left(\gamma_1\,\cos[\Psi]-\gamma_2\,\sin[\Psi]\right)+A'(t)\,\left(\gamma_1\,\sin[\Psi]+\gamma_2\,\cos[\Psi]\right)\Bigg]\\
    \\
    \times\,\left(a_1\,\sin[\gamma_0\,r]+a_2\,\cos[\gamma_0\,r]\right).
  \end{array}
\end{equation}
Thus it is explored that the  metric
\begin{equation}  \label{u41-M-II}
ds^2=-dt^2+A^2(t)\,\Bigg[dr^2+\left(\gamma_1\,\sin\left[\gamma_0\int\dfrac{dt}{A(t)}\right]
+\gamma_2\,\cos\left[\gamma_0\int\dfrac{dt}{A(t)}\right]\right)^2\,\Big(d\theta^2+h^2(\theta)\,d\phi^2\Big)\Bigg],
\end{equation}
admits 6-dimensional conformal Killing vector fields. We can obtain a similar result when $B(t)\,=\,A(t)\,\left(\gamma_1\,\sinh[\Psi]+\gamma_2\,\cosh[\Psi]\right)$,\,\,\,$\Psi\,=\,\gamma_0\,\int\dfrac{dt}{A(t)}$.

\begin{remark} \label{rm-3-0}
It is important to note that the metric (\ref{u41-M-II}) does not admit proper homothetic vector field. When we subtract proper conformal vector fields from CKVFs (\ref{u41-CKVF-2}) by putting $a_1=a_2=0$, we get the four minimum Killing vector fields admitted by Kantowski-Sachs and Bianchi type-III spacetime.
\end{remark}

\textbf{Conformal Killing  Vector Field (4):} In this case the metric functions are obtained as, $B(t)\,=\,A(t)\,\left[\gamma_0+\gamma_1\,\int\dfrac{dt}{A(t)}\right]$, where $\gamma_0$ and $\gamma_1$ are arbitrary non zero constants while $A(t)$ is an arbitrary no where zero function of $t$ only. The components of conformal Killing vector fields are obtained as:
\begin{equation}  \label{u41-CKVF-3}
\left\{
  \begin{array}{ll}
    X^0\,=\,A(t)\,\left[\gamma_0+\gamma_1\,\int\dfrac{dt}{A(t)}\right]\,\left(a_1+2\,\gamma_1\,a_2\,r\right),\\
\\
X^1\,=\,a_{3}+\gamma_1\,a_1\,r+a_2\,\left(\gamma_1^2\,r^2+\left[\gamma_0+\gamma_1\,\int\dfrac{dt}{A(t)}\right]^2\right),\\
\\
X^2\,=a_{4}\,\cos[\phi]+a_{5}\,\sin[\phi],\\
\\
X^3\,=\,a_{6}-\dfrac{h'(\theta)}{h(\theta)}\,\left(a_{4}\,\sin[\phi]-a_{5}\,\cos[\phi]\right),
  \end{array}
\right.
\end{equation}
having conformal factor
\begin{equation}  \label{u41-CF-3}
  \begin{array}{ll}
  \alpha(t,x,y,z)\,=\,\left(\gamma_1+A'(t)\,\left[\gamma_0+\gamma_1\,\int\dfrac{dt}{A(t)}\right]\right)\,\,\left(a_1+2\,\gamma_1\,a_2\,r\right).
  \end{array}
\end{equation}
such that $a_i$, $i=1,...,6$ are arbitrary constants.
Thus, the final form of the conformal Killing vector fields admit  6-dimensional Lie algebra of the metric:
\begin{equation}  \label{u41-M-III}
ds^2=-dt^2+A^2(t)\,\Bigg[dr^2+\left[\gamma_0+\gamma_1\,\int\dfrac{dt}{A(t)}\right]^2\,\Big(d\theta^2+h^2(\theta)\,d\phi^2\Big)\Bigg],
\end{equation}

\begin{remark} \label{rm-4-0}
It is interesting to see that the conformal Killing vector fields ( \ref{u41-CKVF-3}) admit one proper homothetic vector field $t\,\dfrac{\partial}{\partial\,t}$, when $a_2\,=\,0$ and $A(t)\,=\,c_0\,\left(c_1\,t+c_2\right)^{1-c_3/c_1}$, $B(t)\,=\,c_0\,\left(c_1\,t+c_2\right)$, where $c_0\neq0$, $c_1\neq0$, $c_2$ and $c_3$ are arbitrary constants. Five dimensional homothetic vector fields for the above particular metric functions are listed as follows:
\begin{equation}  \label{u41-CKVF-4-00}
\left\{
  \begin{array}{ll}
    X^0\,=\,c_1\,t+c_2,\\
\\
X^1\,=\,c_{3}\,r+c_4,\\
\\
X^2\,=c_{5}\,\cos[\phi]+c_{6}\,\sin[\phi],\\
\\
X^3\,=\,c_{7}-\dfrac{h'(\theta)}{h(\theta)}\,\left(c_{5}\,\sin[\phi]-c_{6}\,\cos[\phi]\right),
  \end{array}
\right.
\end{equation}
where $c_i$, $i=1,...,7$ are arbitrary constants. The conformal factor becomes $\alpha(t,x,y,z)\,=\,c_1$. Again the above homothetic vector fields are in full agreement with \cite{shabbir1}.
\end{remark}
\begin{remark} \label{rm-4-1}
Also, the metric (\ref{u41-M-III}) admits one proper homothetic vector field $t\,\dfrac{\partial}{\partial\,t}+r\,\dfrac{\partial}{\partial\,r}$, when
$a_2\,=\,0$ and $c_1=c_3$ so that the metric funtions become  $A(t)\,=c_0$, $B(t)\,=\,c_0\,\left(c_1\,t+c_2\right)$, where $c_0\neq0$ and $c_1\neq0$. Thus for these particular metric functions, five dimensional homothetic vector fields are obtained as follows:
\begin{equation}  \label{u41-CKVF-4-00}
\left\{
  \begin{array}{ll}
    X^0\,=\,c_1\,t+c_2,\\
\\
X^1\,=\,c_{1}\,r+c_4,\\
\\
X^2\,=c_{5}\,\cos[\phi]+c_{6}\,\sin[\phi],\\
\\
X^3\,=\,c_{7}-\dfrac{h'(\theta)}{h(\theta)}\,\left(c_{5}\,\sin[\phi]-c_{6}\,\cos[\phi]\right),
  \end{array}
\right.
\end{equation}
where $c_i$, $i=1,...,7$ are arbitrary constants. The conformal factor becomes\\
\\ $\alpha(t,x,y,z)\,=\,c_1$. Again the above homothetic vector fields are in full agreement with \cite{shabbir1}.
\end{remark}

\textbf{Conformal Killing  Vector Field (5):} In this case when the metric functions $A(t)$ and $B(t)$ remain arbitrary functions of $t$  other than listed above, then Kantowski-Sach's and Bianchi type III spacetimes admit conformal Killing vector fields as follows:
\begin{equation}  \label{u41-CKVF-5}
\left\{
  \begin{array}{ll}
    X^0\,=\,0,\\
\\
X^1\,=\,a_1,\\
\\
X^2\,=a_{2}\,\cos[\phi]+a_{3}\,\sin[\phi],\\
\\
X^3\,=\,a_{4}-\dfrac{h'(\theta)}{h(\theta)}\,\left(a_{2}\,\sin[\phi]-a_{3}\,\cos[\phi]\right),
  \end{array}
\right.
\end{equation}
where  $a_i$ for $i=1,...,4$ are arbitrary constants and the conformal factor vanishes.

\section{Concircular Vector Fields}

When substituting the above obtained conformal Killing vector fields of different cases in the differential constrains (\ref{2u33-1})--(\ref{8-10u33-2}) turn by turn, we obtain certain restrictions on the metric functions. Solving those restrictions for each case separately, we obtain concircular vector fields for Kantowski-Sachs and Bianchi type-III spacetimes.  The whole procedure is given in the following as different cases:

\textbf{Concircular Vector Fields (1-A):} In this case we reach to the following two restrictions for $A(t)$, that is

$A\,A''-A^{\prime\,2}+\gamma_0^2\,=\,0\,\,\,$ and $\,\,\,A'+\gamma_0\,\coth\left[\Psi\right]\,=\,0$.

The general solution of the above equations together is $A(t)\,=\,\dfrac{\gamma_0\,\sinh[\delta_0\,t+\delta_1]}{\delta_0}$, where $\delta_0\neq0$ and $\delta_1$ are arbitrary constants. Also,\\
$\Psi\,=\,\gamma_0\,\int\dfrac{dt}{A(t)}\,=\,\delta_0\,\int\mathrm{csch}\left[\delta_0\,t+\delta_1\right]\,dt\\
\\
\,=\,-\mathrm{ln}\Big(\mathrm{coth}\left[\delta_0\,t+\delta_1\right]+\mathrm{csch}\left[\delta_0\,t+\delta_1\right]\Big)
\,=\,\mathrm{ln}\Big(\mathrm{coth}\left[\delta_0\,t+\delta_1\right]-\mathrm{csch}\left[\delta_0\,t+\delta_1\right]\Big)$, therefore,

$e^{\Psi}\,=\,\mathrm{coth}\left[\delta_0\,t+\delta_1\right]-\mathrm{csch}\left[\delta_0\,t+\delta_1\right]$,

$e^{-\Psi}\,=\,\dfrac{1}{\mathrm{coth}\left[\delta_0\,t+\delta_1\right]-\mathrm{csch}\left[\delta_0\,t+\delta_1\right]}
\,=\,\mathrm{coth}\left[\delta_0\,t+\delta_1\right]+\mathrm{csch}\left[\delta_0\,t+\delta_1\right]$.

Using the fact: $\mathrm{coth}^2\left[\delta_0\,t+\delta_1\right]-\mathrm{csch}^2\left[\delta_0\,t+\delta_1\right]\,=\,1$ and

$\cosh\left[\Psi\right]\,=\,\dfrac{e^{\Psi}+e^{-\Psi}}{2}\,=\,\mathrm{coth}\left[\delta_0\,t+\delta_1\right]$. The metric function $B(t)$ is obtained as follows:

$B(t)\,=\,\left(\dfrac{A(t)}{\gamma_0}\right)\,\cosh\left[\Psi\right]\,=\,\left(\dfrac{\sinh[\delta_0\,t+\delta_1]}{\delta_0}\right)\,\mathrm{coth}\left[\delta_0\,t+\delta_1\right]\,
=\,\dfrac{\cosh[\delta_0\,t+\delta_1]}{\delta_0}$.
For the above particular metric functions, the components of concircular vector fields become:
\begin{equation}  \label{u41-CVF-1A}
\left\{
  \begin{array}{ll}
    X^0\,=\,\gamma_0\,\Bigg[\left(a_1\,\mathrm{e}^{\gamma_0\,r}+a_2\,\mathrm{e}^{-\gamma_0\,r}\right)\,\cosh\left[\delta_0\,t+\delta_1\right]
    +h'(\theta)\Big[a_3\,\sinh\left[\delta_0\,t+\delta_1\right]+a_4\,\mathrm{e}^{\gamma_0\,r}+a_5\,\mathrm{e}^{-\gamma_0\,r}\Big]\\
    \\
    \,\,\,\,\,\,\,\,\,\,\,\,\,\,\,\,\,\,\,\,\,\,\,\,\,
    +h(\theta)\Bigg(\Big[a_6\,\,\sinh\left[\delta_0\,t+\delta_1\right]+a_8\,\mathrm{e}^{\gamma_0\,r}+a_9\,\mathrm{e}^{-\gamma_0\,r}\Big]\,\cos[\phi]\\
    \\
    \,\,\,\,\,\,\,\,\,\,\,\,\,\,\,\,\,\,\,\,\,\,\,\,\,\,\,\,\,\,\,\,\,\,\,\,\,\,\,\,\,
    \,\,\,\,\,\,\,\,\,\,\,\,\,\,\,\,\,\,\,\,\,\,\,\,\,\,\,\,\,\,\,\,\,\,\,\,\,\,\,\,\,
    +\Big[a_7\,\sinh\left[\delta_0\,t+\delta_1\right]+a_{10}\,\mathrm{e}^{\gamma_0\,r}+a_{11}\,\mathrm{e}^{-\gamma_0\,r}\Big]\,\sin[\phi]\Bigg)\Bigg],\\
\\
X^1\,=\,\delta_0\,\Bigg[a_{12}-\left(a_1\,\mathrm{e}^{\gamma_0\,r}-a_2\,\mathrm{e}^{-\gamma_0\,r}\right)\,\mathrm{csch}\left[\delta_0\,t+\delta_1\right]
-\Bigg(h'(\theta)\left(a_4\,\mathrm{e}^{\gamma_0\,r}-a_5\,\mathrm{e}^{-\gamma_0\,r}\right)\\
\\
+h(\theta)\Big[\left(a_8\,\mathrm{e}^{\gamma_0\,r}-a_9\,\mathrm{e}^{-\gamma_0\,r}\right)\,\cos[\phi]
+\left(a_{10}\,\mathrm{e}^{\gamma_0\,r}-a_{11}\,\mathrm{e}^{-\gamma_0\,r}\right)\,\sin[\phi]\Big]\Bigg)\,\coth\left[\delta_0\,t+\delta_1\right]\Bigg],\\
\\
X^2\,=\,\gamma_0\,\delta_0\,\Bigg[a_{13}\,\cos[\phi]+a_{14}\,\sin[\phi]+h'(\theta)\,\Bigg[\left(a_{6}\,\cos[\phi]+a_{7}\,\sin[\phi]\right)\,\mathrm{sech}\left[\delta_0\,t+\delta_1\right]\\
\\
-\Big[\left(a_8\,\mathrm{e}^{\gamma_0\,r}+a_9\,\mathrm{e}^{-\gamma_0\,r}\right)\,\cos[\phi] +\left(a_{10}\,\mathrm{e}^{\gamma_0\,r}+a_{11}\,\mathrm{e}^{-\gamma_0\,r}\right)\,\sin[\phi]\Big]\,\mathrm{tanh}\left[\delta_0\,t+\delta_1\right]\Bigg]\\
\\
+h(\theta)\,\Big[a_3\,\mathrm{sech}\left[\delta_0\,t+\delta_1\right]-\left(a_4\,\mathrm{e}^{\gamma_0\,r}+a_5\,\mathrm{e}^{-\gamma_0\,r}\right)\,\mathrm{tanh}\left[\delta_0\,t+\delta_1\right]\Big]\Bigg],\\
\\
X^3\,=\,\gamma_0\,\delta_0\,\Bigg[\,a_{15}+\dfrac{h'(\theta)}{h(\theta)}\,\left(a_{14}\,\cos[\phi]-a_{13}\,\sin[\phi]\right)\\
\\
-\dfrac{1}{h(\theta)}\,\Bigg(\left(a_{7}\,\cos[\phi]-a_{6}\,\sin[\phi]\right)\,\mathrm{sech}\left[\delta_0\,t+\delta_1\right]\\
\\
-\Big[\left(a_{10}\,\mathrm{e}^{\gamma_0\,r}+a_{11}\,\mathrm{e}^{-\gamma_0\,r}\right)\,\cos[\phi] -
+\left(a_{8}\,\mathrm{e}^{\gamma_0\,r}+a_{9}\,\mathrm{e}^{-\gamma_0\,r}\right)\,\sin[\phi]\Big]\,\mathrm{tanh}\left[\delta_0\,t+\delta_1\right]\Bigg)\Bigg],
  \end{array}
\right.
\end{equation}
where $a_i$, $i=1,...,15$ are arbitrary constants. Also the conformal factors take the forms:
\begin{equation}  \label{u41-CVF-IA}
  \begin{array}{ll}
    \alpha(t,x,y,z)\,=\,\gamma_0\,\delta_0\,\Bigg[\left(a_1\,\mathrm{e}^{\gamma_0\,r}+a_2\,\mathrm{e}^{-\gamma_0\,r}\right)\,\mathrm{sinh}\left[\delta_0\,t+\delta_1\right]\\
    \\
    +\Big[a_3\,h'(\theta)+\left(a_6\,\cos[\phi]+a_7\,\sin[\phi]\right)h(\theta)\Big]\,\mathrm{cosh}\left[\delta_0\,t+\delta_1\right]\bigg].
  \end{array}
\end{equation}
\begin{equation}  \label{u41-CF-IA}
\beta(t,x,y,z)\,=\,-3\,\delta_0^2\,\alpha(t,x,y,z).
\end{equation}
Therefore we have the following theorem,
\begin{theorem} \label{rm-IA-1}
The Kantowski-Sachs and Bianchi type-III metric
\begin{equation}  \label{u31-M-IA}
ds^2=-dt^2+\delta_0^{-2}\Bigg[\gamma_0^2\,\mathrm{sinh}^2\left[\delta_0\,t+\delta_1\right]\,dr^2+
\mathrm{cosh}^2\left[\delta_0\,t+\delta_1\right]\,\Big(d\theta^2+h^2(\theta)\,d\phi^2\Big)\Bigg],
\end{equation}
where $\gamma_0$ and $\delta_0$ are arbitrary nonzero constants, admits fifteen dimensional concircular vector fields (\ref{u41-CVF-1A}) along with conformal factors (\ref{u41-CVF-IA}) and (\ref{u41-CF-IA}). For the line element (\ref{u31-M-IA}), the relation between Ricci tensor and metric tensor is:
\begin{equation}  \label{u41-M01-1}
  \begin{array}{ll}
R_{ii}\,=\,-3\,\delta_0^2\,g_{ii},\,\,\,\,\,\forall\,\,\,\,\,i\,=\,0,1,2,3.
  \end{array}
\end{equation}
\end{theorem}
This metric is Einstein space and the conformal Killing vector field $X$ agrees with Lemma (\ref{lm-2}). It is easy to satisfy the important relation $\beta\,=\,K\,\alpha$ for $K\,=\,-3\,\delta_0^2$. The above obtained concircular vector fields in Eq. (\ref{u41-CVF-1A}) indicate that the 15-dimensional concircular vector fields of the above spacetime are also the conformal Ricci collineation for the same spacetime. That is:
$$
\mathcal{L}_{X}\,R_{ij}\,=\,2\,\beta\,g_{ij}\,=\,2\,(-3\,\delta_0^2\,\alpha)\,g_{ij}\,=\,2\,\alpha\,R_{ij}.
$$

\textbf{Concircular Vector Field  (1-B):} In this case $B(t)\,=\,\dfrac{A(t)}{\gamma_0}\,\cosh\left[\Psi\right]$,\,\,\,$\Psi\,=\,\gamma_0\,\int\dfrac{dt}{A(t)}$, where $\gamma_0\neq0$ is an arbitrary constant and
$A(t)$ is an arbitrary no where zero function of $t$. The components of concircular vector fields are obtained as:
\begin{equation}  \label{u41-CKV-IB}
\left\{
  \begin{array}{ll}
    X^0\,=\,0,\,\,\,\,\,\,\,\,\,\,\,\,\,\,
X^1\,=\,a_1,\\
\\
X^2\,=\,a_2\,\cos[\phi]+a_3\,\sin[\phi],\\
\\
X^3\,=\,a_4+\dfrac{h'(\theta)}{h(\theta)}\,\left(a_3\,\cos[\phi]-a_2\,\sin[\phi]\right),
  \end{array}
\right.
\end{equation}
where $a_i$, $i=1,...,4$ are arbitrary constants and conformal factors vanish.
Thus, we have the following theorem,
\begin{theorem} \label{rm-IB}
The metric (\ref{u41-M-I}), admits four dimensional concircular vector fields (\ref{u41-CKV-IB}), with vanishing conformal factors.
\end{theorem}
Note that the metric (\ref{u41-M-I}) was admitting fifteen dimensional conformal Killing vector fields, while it is now admitting only four dimensional concircular vector fields. The vanishing conformal factors indicate that these concircular vector fields are just the minimum Killing vector fields admitted by Kantowski-Sach's and Bianchi type-III spacetime.

\textbf{Concircular Vector Fields (2):} In this case the metric functions become:\\ $A(t)\,=\,\delta_1\,\mathrm{e}^{\delta_0\,t}$, $B(t)\,=\,\delta_2\,\mathrm{e}^{\delta_0\,t}$ where $\delta_0$, $\delta_1$ and $\delta_2$ are arbitrary non zero constants such that $\delta_1\neq\delta_2$. For these particular metric functions the components of concircular vector fields are obtained as:
\begin{equation}  \label{u41-CVF-4A}
\left\{
  \begin{array}{ll}
    X^0\,=\,\left(a_1+a_2\,r\right)\,\mathrm{e}^{\delta_0\,t},\\
\\
X^1\,=\,a_3-\delta_0^{-1}\,\delta_1^{-2}\,a_2\,\,\mathrm{e}^{-\delta_0\,t},\\
\\
X^2\,=\,a_4\,\cos[\phi]+a_5\,\sin[\phi],\\
\\
X^3\,=\,a_6+\dfrac{h'(\theta)}{h(\theta)}\,\left(a_5\,\cos[\phi]-a_4\,\sin[\phi]\right),
  \end{array}
\right.
\end{equation}
where $a_i$, $i=1,...,6$ are arbitrary constants. The conformal factors become
\begin{equation}  \label{u41-CF-4A}
  \begin{array}{ll}
    \alpha(t,x,y,z)\,=\,\delta_0\,\left(a_1+a_2\,r\right)\,\mathrm{e}^{\delta_0\,t},
  \end{array}
\end{equation}
\begin{equation}  \label{u41-CF-5A}
\beta(t,x,y,z)\,=\,-3\,\delta_0^2\,\alpha(t,x,y,z).
\end{equation}
The obtained result can be stated in the form of a theorem as follows:
\begin{theorem} \label{rm-IA-1}
The metric
\begin{equation}  \label{u31-M-4A}
ds^2=-dt^2+\mathrm{e}^{2\,\delta_0\,t}\,\Big[\delta_1^{2}\,dr^2+\delta_2^{2}\,\Big(d\theta^2+h^2[\theta]\,d\phi^2\Big)\Big],
\end{equation}
where $\delta_0$, $\delta_1$ and $\delta_2$ are arbitrary non zero constants such that $\delta_1\neq\delta_2$, admits six dimensional concircular vector fields (\ref{u41-CVF-4A}), having conformal factors (\ref{u41-CF-4A}) and (\ref{u41-CF-5A}).
\end{theorem}
For the line element (\ref{u31-M-4A}), the relation between the Ricci tensor and metric tesor are:
\begin{equation}  \label{u41-M01-4A}
  \begin{array}{ll}
R_{ii}\,=\,-3\,\delta_0^2\,g_{ii},\,\,\,\,\,\forall\,\,\,\,\,i\,=\,0,1,2,3.
  \end{array}
\end{equation}
The above obtained concircular vector fields (\ref{u41-CVF-4A}) indicate that the 6-dimensional concircular vector fields of the above spacetime are also the conformal Ricci collineation for the same spacetime. That is:
$$
\mathcal{L}_{X}\,R_{ij}\,=\,2\,\beta\,g_{ij}\,=\,2\,(-3\,\delta_0^2\,\alpha)\,g_{ij}\,=\,2\,\alpha\,R_{ij}.
$$

\textbf{Concircular Vector Field  (3):} When the metric functions $A(t)$ and $B(t)$ take any one of the form $B(t)\,=\,\gamma_0\,A(t)$, $B(t)\,=\,A(t)\,\left(\gamma_1\,\sin[\Psi]+\gamma_2\,\cos[\Psi]\right)$,\,\,\,$\Psi\,=\,\gamma_0\,\int\dfrac{dt}{A(t)}$ or $B(t)\,=\,A(t)\,\left[\gamma_0+\gamma_1\,\int\dfrac{dt}{A(t)}\right]$,  where $\gamma_0$, $\gamma_1$ and $\gamma_2$ are arbitrary non zero constants.
 The components of concircular vector fields are listed below:
\begin{equation}  \label{u41-CKV-4B}
\left\{
  \begin{array}{ll}
    X^0\,=\,0,\,\,\,\,\,\,\,\,\,\,\,\,\,\,
X^1\,=\,a_1,\\
\\
X^2\,=\,a_2\,\cos[\phi]+a_3\,\sin[\phi],\\
\\
X^3\,=\,a_4+\dfrac{h'(\theta)}{h(\theta)}\,\left(a_3\,\cos[\phi]-a_2\,\sin[\phi]\right),
  \end{array}
\right.
\end{equation}
where $a_i$, $i=1,...,4$ are arbitrary constants. The conformal factors vanish, i.e.
\begin{equation}  \label{u41-CF-4B}
  \begin{array}{ll}
    \beta(t,x,y,z)\,=\,\alpha(t,x,y,z)\,=\,0.
  \end{array}
\end{equation}
Thus we have the following theorem:
\begin{theorem} \label{rm-4B}
The metric (\ref{u41-M-4}), (\ref{u41-M-II}) and (\ref{u41-M-III}) admit four dimensional concircular vector fields (\ref{u41-CKV-4B}) whose conformal factors are zero.
\end{theorem}

\section{Summary of the Work}

In this paper,  we investigated concircular vector fields for Kantowski-sach and Bianchi type-III spacetimes. In order to obtain the concircular vector fields components and their conformal factors, we divided this paper into different sections. In $\textbf{(Section 2)}$, we obtained conformal Killing equations and concircular equations. As a first step, we applied some algebraic techniques to reach the general form of conformal Killing vector fields components and the integrability constraints for the complete solution of conformal Killing equations. Those CKVFs components are then utilized into the concircular equations to obtain the integrability conditions for concircular vector fields.\\
 In $\textbf{(Section 3)}$, integrability constraints of conformal Killing equations are solved completely and conformal Killing vector fields along with their conformal factors are obtained. Solutions of integrability constraints of conformal Killing equations, placed different restrictions on the metric function $B(t)$. In most of the cases where Kantowski-Sachs and Bianchi type-III spacetimes admit proper conformal Killing vector fields, $B(t)$ is dependent upon $A(t)$. Each restriction is solved as different case and complete solutions of conformal Killing equations are obtained. In first case it is shown that the metric $(\ref{u41-M-I})$ admits fifteen CKVFs $(\ref{u41-CKVF-1})$. It is also shown that when
$A(t)\,=\,\dfrac{\gamma_0\,\sinh[\delta_0\,t]}{\delta_0}$ and $B(t)\,=\,\dfrac{\cosh[\delta_0\,t]}{\delta_0}$, where $\gamma_0$ and $\delta_0$ are arbitrary non zero constants, spacetime $(\ref{u41-M-I})$ does not admit proper homothetic vector field. In the second case it is shown that when $B(t)\,=\,\gamma_0\,A(t),\,\,\,\,\,\,\gamma_0\neq0$, Kantowski-Sachs and Bianchi type-III spacetime admit six dimensional CKVFs (\ref{u41-CKVF-4}). It is also shown that when $A(t)\,=\,t+c_2$ and $B(t)\,=\,\gamma_0\,\left(t+c_2\right)$,  $\gamma_0\neq0$, the spacetime metric admit one proper homothetic vector field which is in full agreement with the already obtained result of \cite{shabbir1}. In the third case, it is shown that when $B(t)\,=\,A(t)\,\left(\gamma_1\,\sin[\Psi]+\gamma_2\,\cos[\Psi]\right)$,
$\Psi\,=\,\gamma_0\,\int\dfrac{dt}{A(t)},\,\,\,\,\,\,\gamma_0\,\neq\,0$, the spacetime metric (\ref{u41-M-II}) admits six dimensional CKVFs (\ref{u41-CKVF-2}). Moreover, metric in this case admits two proper conformal Killing vector fields and no proper homothetic vector field exists. In the next case four, it is shown that when $B(t)\,=\,A(t)\,\left[\gamma_0+\gamma_1\,\int\dfrac{dt}{A(t)}\right]$,  $\gamma_0,\gamma_1\neq0$, the spacetime metric (\ref{u41-M-III}) admits six dimensional CKVFs (\ref{u41-CKVF-3}). Proper homothtetic vector fields are obtained for two different sets of metric functions. In the first sub case for $A(t)\,=\,c_0\,\left(c_1\,t+c_2\right)^{1-c_3/c_1}$, $B(t)\,=\,c_0\,\left(c_1\,t+c_2\right)$, where $c_0\neq0$, the proper HVF is obtained as $t\,\dfrac{\partial}{\partial\,t}$. In the second sub case for $A(t)\,=c_0$, $B(t)\,=\,c_0\,\left(c_1\,t+c_2\right)$, where $c_0\neq0$ and $c_1\neq0$, the proper HVF is obtained as $t\,\dfrac{\partial}{\partial\,t}+r\,\dfrac{\partial}{\partial\,r}$. It is important to note that both these proper HVFs are in full agreement with the already obtained proper HVFs of G. Shabbir and F. Iqbal \cite{shabbir1}. In the last case of this section it is shown that when $A(t)$ and $B(t)$ are arbitrary no where zero independent functions of $t$ only, then the spacetime under consideration does not admit proper conformal Killing vector field and CKVFs are just the basic four Killing vector fields.\\
In $\textbf{(Section 4)}$, all the results obtained in previous section were put into differential constraints of concircular vector fields, turn by turn. All the results obtained in this section are listed in the form of theorems. In $\textbf{(Case (I-A))}$, concircular vector fields (Eq.\ref{u41-CVF-1A}) are obtained for particular choice of the metric functions. These CVFs correspond to the Einstein space (\ref{u31-M-IA}). It is established that every CKVF is also a CVF (Lemma\ref{lm-2}). Interestingly, the obtained CVFs are also conformal Ricci collineation $(\mathcal{L}_{X}\,R_{ij}\,=\,2\,\alpha\,R_{ij})$. In $\textbf{(Case (I-B))}$, when a relation between metric functions $A(t)$ and $B(t)$ exist as $B(t)\,=\,\dfrac{A(t)}{\gamma_0}\,\cosh\left[\Psi\right]$, where $\gamma_0\neq0$ $\Psi\,=\,\gamma_0\,\int\dfrac{dt}{A(t)}$ such that $A(t)$ is an arbitrary no where zero function of $t$ only then CVFs are just the minimum four KVFs admitted by Kantowski-Sachs and Bianchi type-III spacetime. In $\textbf{(Case 2)}$, six dimensional CVFs are obtained $(Eq.\ref{u41-CVF-4A})$ for special choices of the metric functions. The metric which admits these six dimensional CVFs is given in Eq.$(\ref{u31-M-4A})$. It is also shown that the obtained CVFs in this are conformal Ricci collineation for the same metric. In the last $\textbf{(Case 3)}$, it is shown that when Kantowski-Sachs and Bianchi type-III spacetime have any of the form (\ref{u41-M-4}), (\ref{u41-M-II}) or (\ref{u41-M-III}), the CVFs are just the minimum four KVFs.


\end{document}